\documentclass[aps, pra, amsfonts, amssymb, amsmath, twocolumn]{revtex4-1}

\pdfoutput=1 

\usepackage{amsfonts}
\usepackage{amssymb}
\usepackage{amsmath}
\usepackage{hyperref}           
\hypersetup{colorlinks = True, 
            allcolors = {blue}
}
\usepackage{mathtools}
\usepackage{natbib}
\usepackage{relsize}            
\usepackage{qcircuit}           
\usepackage{graphicx}                  
\usepackage[caption=false]{subfig}     
\usepackage{soul}
\usepackage{multirow}           
\usepackage{bm} 
\usepackage[ruled]{algorithm2e} 

\newcommand{\ket}[1]{ |{#1}\rangle }

\def\rVac{ |\text{vac} \rangle }
\def\rsVac{ | \mathbf{0} \rangle }

\def\rHF{|\text{SD}\rangle}

\def\rAGP{|\text{AGP}\rangle}

\def\rqAGP{|\text{qAGP}\rangle}

\def\rtAGP{|\widetilde{\text{AGP}}\rangle}

\def\rqBCS{|\text{qBCS}\rangle}

\newcommand{\ESP}[2]{S^{#1}_{#2}}
\newcommand{\opESP}[2]{\hat{S}^{#1}_{#2}}
\newcommand{\rESP}[2]{|S^{#1}_{#2} \rangle}

\newcommand{\rDicke}[2]{| D^{#1}_{#2} \rangle}

\newcommand{\Uni}[2]{U_{#1 #2}}
\newcommand{\SCS}[2]{SCS_{#1 #2}}
\newcommand{\Ctwo}[1]{V_{#1 1}}
\newcommand{\Cthree}[2]{V_{#1 #2}}
\newcommand{\Nfac}[4]{\sqrt{\frac{S^{#1}_{#2}}{S^{#3}_{#4}}}}

\newcommand{\BTP}[2]{B^{#1}_{#2}}
\newcommand{\rBTS}[2]{|B^{#1}_{#2} \rangle}

\newcommand{\NfacBTP}[4]{\sqrt{\frac{B^{#1}_{#2}}{B^{#3}_{#4}}}}

\def\rOn{|1\rangle}
\def\rOff{|0\rangle}

\newcommand{\OnP}[1]{ |1\rangle^{\otimes #1} }
\newcommand{\OffP}[1]{ |0\rangle^{\otimes #1} }

\newcommand{\cdag}[1]{ {c}^{\dagger}_{#1} }
\newcommand{\cn}[1]{{c}_{#1}}

\def\GamD{ \mathlarger{{{\Gamma}}}^{\dagger} }

\newcommand{\Pdag}[1]{\mathbf{P}^{\dagger}_{#1} }
\newcommand{\N}[1]{ \mathbf{N}_{#1} }
\newcommand{\Ntot}[1]{\mathbf{N}}
\newcommand{\Pp}[1]{ \mathbf{P}_{#1} }

\newcommand{\Jst}[1]{\boldsymbol{J}_{#1} }

\newcommand{\Sx}[1]{ {X}_{#1} }
\newcommand{\Sy}[1]{ {Y}_{#1} }
\newcommand{\Sz}[1]{ {Z}_{#1} }
\newcommand{\Su}[1]{ \sigma^{+}_{#1} }

\newcommand{\Eesp}[2]{\langle {S}^{#1}_{#2} | {S}^{#1}_{#2} \rangle}

\newcommand{\SumOp}[2]{ \sum_{1 \leq p_1 < \cdots < p_{#2} \leq #1} }
\newcommand{\icol}[1]
    {\left(\begin{smallmatrix}#1\end{smallmatrix}\right)}

\newcommand{\bigO}[1]{\mathcal{O}(#1)}

\newcommand{\Ham}{H} 

\newcommand{\Eq}[1]{Eq.~({#1})}
\newcommand{\Sec}[1]{Sec.~{#1}}
\newcommand{\Fig}[1]{Fig.~{(#1)}}
\newcommand{\Alg}[1]{Algorithm~{#1}}
\newcommand{\Reference}[1]{Ref.~{#1}}
\newcommand{\Appx}[1]{Appendix~{#1}}

\def\HermConj{\text{h.c.}}

\begin{document}

\title{
  State preparation of AGP on a quantum computer without number projection
}

\author{Armin Khamoshi}
    \affiliation{Department of Physics and Astronomy, Rice University, Houston, Texas 77005, USA}

\author{Rishab Dutta}
    \affiliation{Department of Chemistry, Rice University, Houston, Texas 77005, USA}

\author{Gustavo E. Scuseria}
    \affiliation{Department of Chemistry, Rice University, Houston, Texas 77005, USA}
    \affiliation{Department of Physics and Astronomy, Rice University, Houston, Texas 77005, USA}

\date{\today}


\begin{abstract} 

The antisymmetrized geminal power (AGP) is equivalent to the number projected Bardeen--Cooper--Schrieffer (PBCS) wavefunction. 
It is also an elementary symmetric polynomial (ESP) state. 
We generalize previous research on deterministically implementing the Dicke state to a state preparation algorithm for an ESP state, or equivalently AGP, on a quantum computer. 
Our method is deterministic and has polynomial cost, and it does not rely on number symmetry breaking and restoration. 
We also show that our circuit is equivalent to a disentangled unitary paired coupled cluster operator and a layer of unitary Jastrow operator acting on a single Slater determinant. 
The method presented herein highlights the ability of disentangled unitary coupled cluster to capture non--trivial entanglement properties that are hardly accessible with traditional Hartree--Fock based electronic structure methods.

\end{abstract}

\maketitle


\section{Introduction} \label{sec:intro}

State preparation is a crucial part of any quantum algorithm. 
For variational quantum algorithms in physics and chemistry, preparation of the ansatz usually starts from a single Slater determinant (SD), such as the Hartree-Fock (HF) wavefunction. 
Preparing HF is relatively straightforward and inexpensive, even on noisy intermediate-scale quantum (NISQ) era devices 
\cite{QCQC2019,QCQC2020,QCHF2020}. 
Generally, any single Slater determinant can be prepared by applying the Thouless theorem using a series of Givens rotations on a quantum computer \cite{Wacker2015,Kivlichan2018,Boixo2018,QCHF2020}. 
Preparation of the HF state is the starting point for many correlated methods, including variants of unitary coupled-cluster theory (uCC) on a quantum computer which is an active area of research \cite{AAGUCC2022}.

State preparation, however, may not be as straightforward when the initial state is not a single Slater determinant. 
Examples of such states include Bethe ansatz \cite{Economou2021,Economou2022,Nepomechie2022}, and geminal--based wavefunctions \cite{AGPQC2021,AGPQC2022,eAGPQC2022}, which have a long history in physics and chemistry. 
Antisymmetrized geminal power (AGP), also known as the projected Bardeen--Cooper--Schrieffer (PBCS) wavefunction \cite{RingBook}, is perhaps the simplest of all geminal--based wavefunctions. 
Recent research has shown that
AGP is a promising initial state for correlated methods, especially for the strong correlation regime, including molecules and model Hamiltonians such as pairing and Fermi--Hubbard
\cite{Neuscamman2013,Neuscamman2016,Sorella2014,Sorella2019,TomAGPCI2019,TomAGPRPA2020,GRAGP2020,Kurashige2020,AGPCC2021,LCAGP2021,AGPQC2021,AGPQC2022}. 
Some of the authors have developed unitary correlated methods based on AGP within a variational quantum eigensolver (VQE) approach \cite{AGPQC2021,AGPQC2022}.
Nevertheless, it has been challenging to find a deterministic and efficient circuit that prepares AGP exactly on a quantum computer. 
The reason for this stems in part from the fact that AGP is a superposition of ${M \choose N}$ number of states, where $M$ and $N$ are the number of paired orbitals and electron pairs respectively; as we shall see later, the amplitude of every state uniquely corresponds to the product of one of $N$-combinations of a set of geminal coefficients $\{\eta_1, \hdots \eta_M\}$ \cite{AGPRDM2019}. 
As such, some of the more general state preparation algorithms \cite{Salomaa2005,Plesch2011,Araujo2021,LowDepth2021}
scale exponentially in the number of qubits or depth for implementing AGP on a quantum computer.

To overcome the challenges of efficiently computing overlaps over AGP on a quantum computer, some of the authors of this paper have proposed a symmetry breaking and restoration approach \cite{AGPQC2021,AGPQC2022}.
Since AGP is formally equivalent to PBCS, we can prepare the BCS state efficiently and project the overlaps via gauge integration \cite{AGPQC2021}, phase estimation \cite{SymQC2020}, or post--selection \cite{AGPQC2022}. 
Symmetry--projection is intrinsically a non-unitary operation \cite{RingBook}. 
Thus, the aforementioned methods necessarily rely on ancilla qubits or collapse of the wavefunction to project number symmetry. 
While for all practical purposes, these are workable and efficient solutions, we would like to propose an alternative method in this paper. 

We present a unitary circuit that implements AGP on a quantum computer without number projection. 
The key idea is that the AGP wavefunction can be written as an \textit{elementary symmetry polynomial} (ESP) state wherein the wavefunction obeys a special recursive decomposition formula \cite{AGPRDM2019,LCAGP2021}. 
This allows us to devise a divide--and--conquer algorithm that evolves a single Slater determinant state into AGP at polynomial cost. 
Our implementation follows the work of \Reference{\cite{BartschiDicke2019}}, which introduced a deterministic algorithm to implement the Dicke state on a quantum computer. 
As we shall see, the Dicke state is a special case of AGP where all geminal coefficients take the same value. This is known as \textit{extreme} AGP in quantum chemistry \cite{Coleman1965}. 
Thus, our method can be viewed as generalizing the preparation of the Dicke state onto preparing ESP states on a quantum computer. 
That is, instead of having all determinants share the same amplitude as in the Dicke state, we make the amplitude of each determinant uniquely correspond to a term in a given elementary symmetric polynomial. 

We also show that our circuit is equivalent to a disentangled form of the unitary paired coupled cluster (upCC) with generalized doubles and quadruples 
\cite{AyersJCTC2013,Stein2014,SenCC2014,ExactUCC2019,gpUCC2019,Elfving2021,Kottmann2022,GooglePair2022}, and a unitary one--body Jastrow part acting on an initial SD state. 
There has been much discussion in the literature about the accuracy and advantages of the disentangled unitary coupled cluster ansatz \cite{ExactUCC2019,ADAPT2019,Grimsley2020,AAGUCC2022}. 
This work presents one advantage of the single--reference based disentangled upCC compared to the traditional CC methods. 
The traditional methods often break down in the strongly correlated regime, especially when the many--body interactions are attractive \cite{BCSCC2014,pECC2015,PoST2016,TomAGPRPA2020}. 
In such cases, the remedy is often to allow the particle--number symmetry to break at the mean--field and later restore it \cite{TomAGPRPA2020,PQT2011,PHFBCC2019,Sheikh2021}.
A prime example is the pairing Hamiltonian for which neither symmetry--adapted nor symmetry--broken single reference CC method accurately captures the ground state energy and the superfluid phase transition \cite{BCSCC2014, PoST2016, TomAGPRPA2020}; whereas, the number symmetry restored wavefunction, namely AGP, is well--behaved in all regimes and approaches the correct limit in both weakly and strongly correlated regimes \cite{TomAGPRPA2020}. 
As such, and in contrast to traditional CC, we show that there exists a particular ordering of the disentangled uCC based on a single Slater determinant that contains AGP without breaking number symmetry.

From the viewpoint of symmetry breaking and restoration methods on a quantum computer, our algorithm provides a practical advantage in having fewer measurements compared to number projection. 
This is particularly advantageous for cases in which we want to deliberately break and restore multiple symmetries atop of AGP \cite{AGPQC2022}, e.g. spin angular momentum, since the measurement cost of restoring additional symmetries could easily mount. 
In essence, our method trades having a deterministic algorithm to prepare a number--symmetry restored wavefunction in exchange for a deeper circuit at $\bigO{M^2}$ cost. 

It is noteworthy that all reduced density matrices (RDMs) over AGP can be computed efficiently on a classical computer, and the AGP wavefunction itself can be optimized at mean--field cost \cite{PQT2011,AGPRDM2019,LCAGP2021}.
The advantage of implementing AGP on a quantum computer transpires for post-AGP correlated methods, such as variants of unitary coupled cluster theory, that are costly on classical computers but can be done efficiently on a quantum computer 
\cite{AGPQC2021,Kurashige2020,AGPQC2022}.
Similarly, the ESP state preparation introduced in this paper could be applicable as the first step for more sophisticated algorithms in quantum information theory \cite{Childs2002,Farhi2014,Hadfield2019}.

The rest of this paper is organized as follows: In \Sec{\ref{sec:preliminary}}, we discuss the ESP state and its equivalence with AGP. 
\Sec{\ref{subsec:agp_prep}} presents a deterministic preparation scheme for AGP in the seniority--zero implementation before extending it for seniority nonzero systems in \Sec{\ref{subsec:nzero_seniority}}.
Finally, we conclude with a discussion in \Sec{\ref{sec:discussions}}. 


\section{Preliminaries} \label{sec:preliminary}

\subsection{Elementary symmetric polynomial states}\label{sec:esp_state}

An elementary symmetric polynomial (ESP) of degree $n$ in $m$ variables, 
$ \mathbf{x} = \{ x_1, x_2, \cdots, x_m \} $, can be defined as
\begin{equation} \label{eq:esp}
  \ESP{m}{n} \left( \mathbf{x} \right)
  = \SumOp{m}{n} x_{p_1} \cdots x_{p_n}, \quad 1 \leq n \leq m,
\end{equation}
where 
$ \ESP{m}{0} \left( \mathbf{x} \right) = 1 $, and 
$ \ESP{m}{n} \left( \mathbf{x} \right) = 0 $ for $n > m$. 
The right-hand side is a linear combination of all ${m \choose n}$ distinct ways we can pick $n$ variables from $\mathbf{x}$ and multiply them. 
ESPs are building blocks of symmetric polynomials and feature remarkable analytic properties \cite{MacdonaldBook,NelsonBook,Harwell1996}. 
Of main interest to this paper is the following recursion formula \cite{SumESP1974}
\begin{equation} \label{eq: recesp}
  \ESP{m}{n} \left( \mathbf{x} \right)
  = x_p \: \ESP{m - 1}{n - 1} \left( \mathbf{x} \backslash x_p \right)
  + \ESP{m - 1}{n} \left( \mathbf{x}\backslash x_p \right),
\end{equation}
where $\mathbf{x}\backslash x_p$ represents exclusion of $x_p$ from the full set \textbf{x}.
\Eq{\ref{eq: recesp}} says any $\ESP{m}{n} \left( \mathbf{x} \right)$ can be decomposed as a sum of two ESPs---one in which all summands contain the arbitrary variable $x_p$, 
i.e. 
$ x_p \: \ESP{m - 1}{n - 1} \left(\mathbf{x}  \backslash  x_p \right) $, 
and one that excludes it, i.e.
$ \ESP{m - 1}{n} \left( \mathbf{x}  \backslash  x_p \right) $. 
\Eq{\ref{eq: recesp}} is key to computing ESPs efficiently via a divide-and-conquer algorithm with a binary tree structure \cite{SumESP1974,Ipsen2011,Jiang2016}.
Hereafter, we drop the parentheses in the ESP expression to avoid a profusion of variables and indices in later sections. 

While in many instances 
$ \mathbf{x} = \{ x_1, \cdots, x_m \} $ is assumed to be a set of complex numbers, one can envision an ESP over a set of operators as well \cite{LCAGP2021}. 
For example, consider the set of Pauli raising operators in an $m$-qubit system, 
$ \{\Su{p}\}_{p=1}^{m} $, where 
$ \Su{p} \equiv (\Sx{p} - i \Sy{p})/2 $, and 
$X_p$ and $Y_p$ are the standard Pauli matrices acting on qubit $p$. 
Then, we can let 
\begin{equation} \label{eq:esp_operator}
  \opESP{m}{n}
  = \SumOp{m}{n} \left(x_{p_1}\Su{p_1}\right) \cdots \left( x_{p_n} \Su{p_n} \right)
\end{equation}
to be an ESP of operators over the set $\{ x_1 \Su{1}, \cdots, x_m \Su{m} \}$, where we put a hat on the left-hand side of \Eq{\ref{eq:esp_operator}} to distinguish it from the ESP of scalars \Eq{\ref{eq:esp}}. 

We now define the ESP state. 
Let $m$ be the number of qubits and let $n$ denote the number of individual qubits in the $\ket{1} = \icol{0\\1}$ state. 
Then, the ESP state of degree $n$ over $m$ qubits with coefficients 
$ \{x_1, \cdots, x_m \} $, 
$\rESP{m}{n}$, can be defined as 
\begin{multline} \label{eq:ESPstate}
    \rESP{m}{n} \equiv 
    \frac{1}{\sqrt{\ESP{m}{n}}} 
    \SumOp{m}{n} x_{p_1} \cdots x_{p_n} \\
    \ket{0\cdots1_{p_n} \cdots 1_{p_2}\cdots 1_{p_1} \cdots 0}, 
\end{multline}
where $1/\sqrt{\ESP{m}{n}}$ is the normalization factor derived from
\begin{equation} \label{eq: espnorm}
  \ESP{m}{n}
  = \Eesp{m}{n}
  = \SumOp{m}{n} |x_{p_1}|^2 \cdots |x_{p_n}|^2.
\end{equation}
To put it differently, if we let $\opESP{m}{n}$ be the operator defined in \Eq{\ref{eq:esp_operator}}, then the ESP state is equivalent to
\begin{equation} \label{eq:ESPstate_from_operator}
    \rESP{m}{n} = \frac{\opESP{m}{n}}{\sqrt{\ESP{m}{n}}} \: \rsVac{},
\end{equation}
where we define $\rsVac{} \equiv \ket{0}^{\otimes m}$.

Similar to Eq.~\eqref{eq: recesp}, we can recursively split the ESP state, $\rESP{m}{n}$, into a superposition of two orthogonal states. For a given $p,q$ such that 
$ 1 \leq q  \leq n < p \leq m $, we can write
\begin{align} \label{eq: recagp}
    \rESP{p}{q}
  &= \Nfac{p-1}{q-1}{p}{q} \: x_{m - p + 1} \: \rESP{p-1}{q-1} \rOn \nonumber 
  \\
  &+ \Nfac{p-1}{q}{p}{q} \: \rESP{p-1}{q} \rOff,
\end{align}
where we have used the shorthand notation $\ket{x}\ket{y}$ to represent $\ket{x}\otimes \ket{y}$.
Here, in every recursion step of \Eq{\ref{eq: recagp}}, we choose to do the splitting at the rightmost qubit for brevity, and for other reasons that will become apparent in future sections. It should be noted that the splitting itself can be done at any arbitrary qubit in a manner analogous to \Eq{\ref{eq: recesp}}.

Having defined a general ESP state, we highlight a special case in which $x_p = 1$ for all $p$. 
The resulting state is equivalent to the Dicke state \cite{Dicke1954}, which we can write as follows 
\begin{multline} \label{eq: dicke}
  \rDicke{m}{n}
  = \sqrt{ \frac{1}{{m \choose n}} } \SumOp{m}{n}  
  \\
  \ket{0\cdots1_{p_n} \cdots 1_{p_2}\cdots 1_{p_1} \cdots 0}.
\end{multline}
The recursion relation of the Dicke state, a special case of \Eq{\ref{eq: recagp}}, has been applied to design some of its preparation algorithms \cite{BartschiDicke2019,Mukherjee2020,Aktar2022}.

\subsection{AGP as an ESP state}\label{sec:AGP_as_esp}

We now turn our attention to AGP as an ESP state.
AGP is a geminal--based wavefunction, where a geminal is defined as the wavefunction of electron pairs in quantum chemistry \cite{Coleman1963,Coleman1965,Surjan1999}.
A geminal creation operator in its natural orbital basis can be defined as  
\begin{equation} \label{eq: gem}
  \GamD
  = \sum_{p = 1}^M \: \eta_p \: \Pdag{p},
\end{equation}
where $\Pdag{p} = \cdag{p} \cdag{\bar{p}}$ is the pair creation operator, and $\cdag{p}$, $\cdag{\bar{p}}$ represent fermion creation operators acting on orbitals $p$ and its paired companion $\bar{p}$ respectively \cite{TomAGPRPA2020,AGPQC2022}; $M$ is the total number of paired orbitals.
The pairing operators $\Pdag{p}$, $\N{p}$, $\Pp{p}$ are the generators of an $su(2)$ Lie algebra,
\begin{subequations}\label{eq:su2_pairing}
\begin{align}
\left[ \Pp{p}, \Pdag{q} \right] 
&= \delta_{pq} \left( 1 - \N{q} \right), 
\\
\left[ \N{p}, \Pp{q}^{\dagger} \right] 
&= 2 \delta_{pq} \Pdag{q},
\end{align}
\end{subequations}
where 
$ \N{p} 
= \cdag{p}\cn{p} 
+ \cdag{\bar{p}} \cn{\bar{p}} $ is the pair number operator.
The geminal coefficients $\{ \eta_p \}$ are, in general, complex-valued numbers
\begin{equation} \label{eq:gem_coeff_sep}
  \eta_p 
  = | \eta_p | \: e^{i \alpha_p}, 
\end{equation}
where $| \eta_p |$ is the geminal coefficient magnitude and $e^{i \alpha_p}$ is the corresponding phase.

Neglecting the normalization factor, the AGP wavefunction can be defined as
\begin{equation} \label{eq: gemagp}
  \rAGP
  = \frac{1}{N!} \: \left( \GamD \right)^N \; \rVac, 
\end{equation}
where $N$ is the number of pairs, i.e. $2N$ fermions, and $\rVac$ is the physical vacuum. 
By expanding the expression for the geminal power 
$ \left( \GamD \right)^N / N! $, we get an ESP operator
\begin{equation} \label{eq: gempower}
  \opESP{M}{N} 
  = \SumOp{M}{N} \left(\eta_{p_1}\Pdag{p_1}\right) \cdots \left(\eta_{p_N} \Pdag{p_N}\right),
\end{equation}
of degree $N$ over the set $\{\eta_p \Pdag{p}\}_{p=1}^M$.
Clearly, this implies AGP can be expressed as an ESP state based on \Eq{\ref{eq:esp_operator}}.
The ESP structure of AGP has been applied to develop efficient algorithms for computing AGP expectation values \cite{AGPRDM2019,LCAGP2021} and generate new AGPs from a reference AGP \cite{LCAGP2021}.

AGP encoding on a quantum computer can be achieved in several ways: If one is not interested in breaking the fermion pairs, one could pursue a \textit{seniority--zero} implementation. 
The term \textit{seniority} refers to the number of unpaired fermions in a given pairing scheme \cite{RingBook, Seniority2011}. 
In this case, we can let the qubit states $\rOff_p$ and $\rOn_p$ represent empty and doubly-occupied paired orbital $p$ respectively, and map
\begin{subequations} \label{eqs:pair-qubit-mapping}
\begin{align}
    \rVac    &\mapsto \ket{0}^{\otimes M}, \\
    \Pdag{p} &\mapsto \Su{p},\\
    \N{p}    &\mapsto I - \Sz{p}.
\end{align}
\end{subequations} 
where $I$ and $Z$ are the identity and the standard Pauli $Z$ operator respectively. 
\Eq{\ref{eqs:pair-qubit-mapping}} are collectively known as the paired encoding \cite{AGPQC2021,Elfving2021,AGPQC2022}. 

If we are interested in incorporating broken pair excitations atop of AGP, we can resort to the fermionic encoding. 
This can be achieved via the Jordan-Wigner (JW) transformation \cite{JordanWigner1928} for example, where we let every qubit represent the occupation number of the spin-orbitals $p$ or $\bar{p}$, and map
\begin{subequations} \label{eqs:jw-mapping}
\begin{align}
    \rVac    
    &\mapsto \ket{0}^{\otimes 2M}, 
    \\
    \cdag{p} 
    &\mapsto \Su{p} \: \prod_n \Sz{n}, 
    \\ 
    \cdag{\bar{p}} 
    &\mapsto \Su{\bar{p}} \: \prod_n \Sz{n}, 
\end{align}
\end{subequations} 
where $\prod_n \Sz{n}$ 
represents the corresponding JW strings for each fermionic creation operator. 
This is the same approach used in \Reference{\cite{AGPQC2022}} to correlate AGP via a disentangled unitary coupled cluster method. 

We could also envision a different formulation of AGP in terms of qubits or spin--$1/2$ systems which can be used to treat general Hamiltonians that get mapped to the $su(2)$ algebra. 
We shall elaborate more on this approach in \Sec{\ref{subsec:qubit-AGP}.}


\section{Algorithm} \label{sec:algorithm}

In this section, we generalize the algorithm of \Reference{\cite{BartschiDicke2019}} to prepare any AGP on a quantum computer. 
For clarity, we first formulate the algorithm for the paired encoding implementation in \Sec{\ref{subsec:agp_prep}}. 
We then generalize AGP preparation to nonzero seniority in \Sec{\ref{subsec:nzero_seniority}}.


\subsection{AGP preparation}\label{subsec:agp_prep}


\begin{figure}[t]
\centering
\includegraphics[width=0.9\columnwidth]{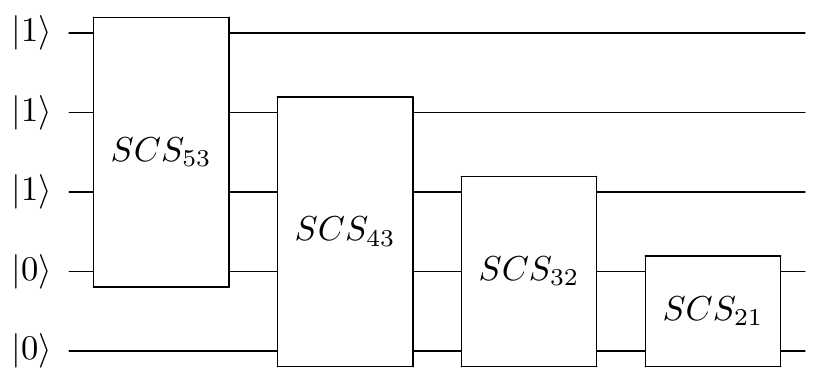}
\caption{Construction of $\Uni{M}{N}$ in terms of the smaller \textit{split and cyclic shift (SCS)} circuits following Eq.~\eqref{eq: unitary}, shown here for $\Uni{5}{3} \: \ket{00111}$ as an example.}
\label{fig: block}
\end{figure}

\begin{figure*}[t]
\centering
\subfloat[][]{
  \includegraphics[width=0.70\columnwidth]{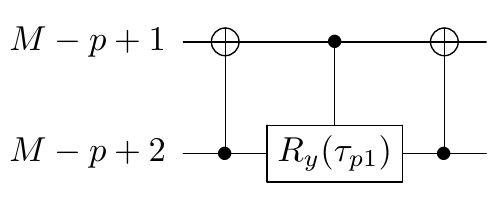}
  \label{fig: two_gates}
}
\hskip10ex
\subfloat[][]{
  \includegraphics[width=0.94\columnwidth]{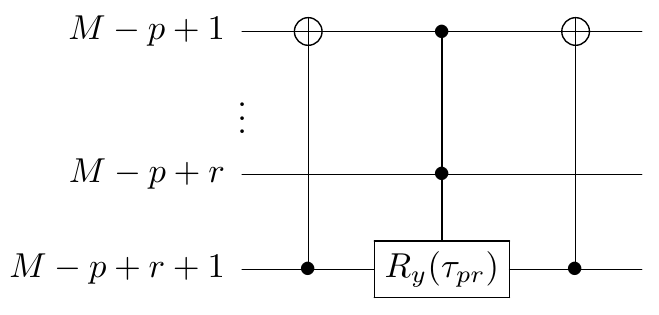}
  \label{fig: three_gates}
}
\caption{
  Circuits for the two-body and three-body generalized excitations from \Eq{\ref{eq:scs_decomp}}.    
  The left and right panels show the construction of the two-qubit gate $\Ctwo{p}$ and the three-qubit gate $\Cthree{p}{r}$ respectively. 
  The scalars inside $R_y$ gates are 
  $\tau_{pr} = 2 \arccos \left( \theta_{pr} \right)$, where $\{ \theta_{pr} \}$ are defined in \Eq{\ref{eq: scs}}.
}
\label{fig: gates}
\end{figure*}

\begin{figure}[b]
\centering
\includegraphics[width=0.78\columnwidth]{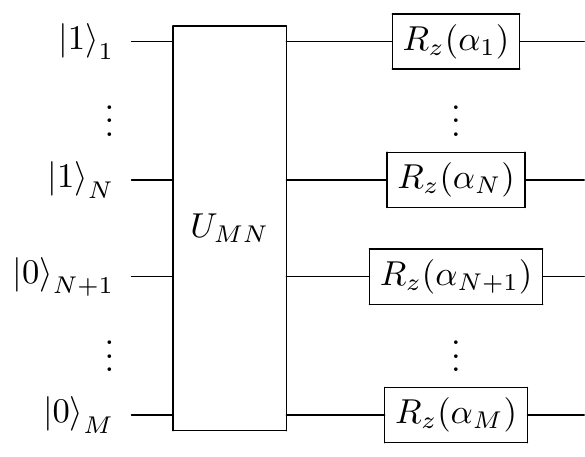}
\caption{
  Complete circuit illustration for preparing the ESP state $\rESP{M}{N}$, which equivalently prepares AGP with $N$ pairs in $M$ orbitals in the paired encoding framwork.
}
\label{fig: fullagp}
\end{figure}

Consider the paired encoding of \Sec{\ref{sec:AGP_as_esp}}.
Our initial state is a SD of fermion pairs
\begin{equation} \label{eq: hf}
  \rHF
  = \OffP{M - N} \OnP{N}.
\end{equation}
Our goal is to create a unitary operator, $\mathcal{U}$, that evolves \Eq{\ref{eq: hf}} into AGP. 
$\mathcal{U}$ is composed of two parts: First, a unitary $\Uni{M}{N}$ that prepares an AGP state with the geminal coefficient magnitudes only; second, a unitary Jastrow operator, $\mathcal{J}$, that adds the corresponding phase of each geminal coefficient. 
Together, 
$ \mathcal{U} = \mathcal{J} \: \Uni{M}{N} $ creates an ESP state of degree $N$ over $M$ qubits with a set of complex--valued coefficients, $\{\eta_p\}$, as desired, 
\begin{equation} \label{eq: evolve}
  \rAGP 
  = \mathcal{U} \: \rHF
  = \mathcal{J} \: \Uni{M}{N} \: \rHF.
\end{equation}
In what follows, we first show how $\Uni{M}{N}$ can be constructed from smaller unitaries, down to elementary gates, following similar steps as in \Reference{\cite{BartschiDicke2019}}. 
Although our proof strategy for this algorithm is by construction,  proofs of \Reference{\cite{BartschiDicke2019}} can be trivially extended to our method by using the splitting property of \Eq{\ref{eq: recagp}} and the unitary rotation angles introduced in this section. 
Later in this section, we turn our attention to the Jastrow operator. 

The building blocks of $\Uni{M}{N}$ are the so-called \textit{split and cyclic shift} ($SCS$) operators \cite{BartschiDicke2019} 
\begin{equation} \label{eq: unitary}
  \Uni{M}{N} 
  = \prod_{p = 2}^N \SCS{p}{p - 1} \prod_{p = N + 1}^M \SCS{p}{N},
\end{equation}
which together implement a bottom-up preparation of an ESP state based on a nested application of \Eq{\ref{eq: recagp}}.
\Fig{\ref{fig: block}} provides a schematic example of how $\Uni{M}{N}$ is constructed in terms of the $SCS$ operators for $M=5$ and $N=3$. 
For a given $1 \leq q \leq N < p \leq M$, we want $\SCS{p}{q}$ to implement \Eq{\ref{eq: recagp}}. 
To do that, we define  
\begin{equation} \label{eq: rotagp}
  \theta_{p q} 
  \equiv | \eta_{M - p + 1} | \: \Nfac{p - 1}{q - 1}{p}{q},
\end{equation}
where the ESPs of the right--hand side are over $\{ |\eta_p |^2 \}$.
We want $\SCS{p}{q}$ to act as follows
\begin{align} \label{eq: scs}
  & \SCS{p}{q} \:
  \OffP{q - r + 1} \OnP{r}
  = \nonumber
  \\ 
  &\theta_{p r} \: \OffP{q - r + 1} \OnP{r} 
  + \sqrt{1 - \theta_{p r}^2} \: \OffP{q - r} \OnP{r} \rOff,
\end{align}
while acting as an identity when encountering other states through \Eq{\ref{eq: unitary}}. 
Put in simple words, when $\SCS{p}{q}$ encounters a state composed of $r$ consecutive number of $\ket{1}$ qubits, say $\ket{011 \cdots 1}$, it splits it into a superposition of two states: The original state with amplitude $\theta_{pr}$ and another state with amplitude $\sqrt{ 1 - \theta_{pr} }$ such that the $r$ consecutive  $\rOn$ qubits have been shifted by exactly one qubit, $\ket{11 \cdots 10}$; that is 
$ \SCS{p}{q} \: \ket{011 \cdots 1} 
= \theta_{pr} \: \ket{011 \cdots 1} 
+ \sqrt{1 - \theta_{pr}^2} \: \ket{11 \cdots 10} $.
When $\SCS{p}{q}$ acts on $\ket{00 \cdots 0}$ or $\ket{11 \cdots 1}$, it acts as an identity. 
A less obvious point is that when $\Uni{M}{N}$ acts on the initial SD state of \Eq{\ref{eq: hf}}, the ordering in \Eq{\ref{eq: unitary}} guarantees each $\SCS{p}{q}$ operator will encounter only the aforementioned states. 

Note that if we let $\eta_p = 1$ for all $p$, we recover the Dicke state rotation angles, $\theta_{p q} = \sqrt{q/p}$, of \Reference{\cite{BartschiDicke2019}}. 

We now discuss how $\SCS{p}{q}$ can be implemented.
Any given $\SCS{p}{q}$ could encounter any of the states containing $r \in \{0, \cdots, q\}$ consecutive $\ket{1}$ qubits. 
Therefore, it must be a product of at most $q$ smaller unitaries that carry out the so-called \textit{split and shift} for each of those possibilities.
An $\SCS{p}{q}$ operator can be written as a product of a two-qubit unitary $\Ctwo{p}$ and a sequence of three-qubit operators $\Cthree{p}{r} \: (1 < r \leq q)$ as follows \cite{BartschiDicke2019}  
\begin{equation} \label{eq:scs_decomp}
  \SCS{p}{q} 
  = \Cthree{p}{q} \cdots \Cthree{p}{1}.
\end{equation}
Following \Eq{\ref{eq: unitary}}, we want $\Ctwo{p}$ to act non--trivially when 
\begin{align}
  \Ctwo{p} \ket{01}  = \theta_{p1} \ket{01} + \sqrt{1 - \theta_{p1}^2} \: \ket{10},  
\end{align}
and as an identity elsewhere, where the indices for the two qubits are $M - p + 1$ and $M - p + 2$ respectively. 
This can be implemented using the well--known Givens rotation circuit as shown in \Fig{\ref{fig: two_gates}}. 
Written in the second quantized language, $\Ctwo{p}$ is nothing but
\begin{align}
    \Ctwo{p} = e^{ \tau_{p1} \: 
    \left( 
        \Pdag{M - p + 2} \Pp{M - p + 1} - \: \HermConj 
    \right) / 2 },
\end{align}
where $\tau_{pq} = 2 \arccos \left( \theta_{pq} \right)$. 
Indeed, this is a doubles excitation in unitary paired coupled cluster theory with general indices 
\cite{gpUCC2019,AGPQC2021,Elfving2021,Kottmann2022,GooglePair2022}, also known as \textit{pair--hoppers} \cite{AGPQC2021,AGPCC2021}.

Similarly, for all $ \Cthree{p}{r} \: (1 < r \leq q) $ we want 
\begin{align}
\Cthree{p}{r} \: \rOff \OnP{r} =
\theta_{pr} \: \rOff \OnP{r} + \sqrt{1 - \theta_{pr}^2} \: \OnP{r}\ket{0},  
\end{align}
and identity otherwise, where the indices of the qubits above run from $M - p + 1$ to $M - p + r + 1$. 
It is sufficient for $\Cthree{p}{r}$ to act on the rightmost and the last two qubits only, as it is guaranteed that the $\ket{1}$ qubits appear contiguously by the construction of \Eq{\ref{eq: unitary}} acting on the initial SD. 
As such, the circuit for $\Cthree{p}{r}$ can be constructed as shown in \Fig{\ref{fig: three_gates}}. 
Although this is a three-qubit circuit, in practice, it can be further decomposed in terms of one- and two-qubit gates only \cite{BartschiDicke2019}. 

Written in the second quantized language, $\Cthree{p}{r}$ corresponds to   
\begin{align} \label{eq:upCC-quadruples}
    \Cthree{p}{r} = e^{ \tau_{pr} \: \left( 
  \Pdag{M - p + r + 1} \N{M - p + r} \Pp{M - p + 1}
  - \: \HermConj \right) / 4 },
\end{align}
where $\tau_{pr} = 2 \arccos \left( \theta_{pr} \right)$ as before.
Since $ \N{p} = 2 \Pdag{p} \Pp{p}$ in the seniority--zero space, this is equivalent to a \text{quadruples} excitation in upCC theory. 

In summary, $\Uni{M}{N}$ acting on an SD prepares an AGP of $N$ pairs in $M$ paired orbitals with the magnitudes of geminal coefficients $\{|\eta_p|\}$. 
Since the building blocks of $\Uni{M}{N}$ are the doubles and quadruples excitations, we have shown a disentangled form of generalized upCC that transforms a single Slater determinant into a AGP state. 
The circuit discussed so far uses $M$ qubits and has a depth that scales as $\bigO{M^2}$. 
On a quantum computer with limited connectivity, the scaling could be different but it is no worse than $\bigO{M^3}$ \cite{Tannu2019}. 
We have analytically worked out a step--by--step example of how $\Uni{M}{N}$ is constructed and acts on the initial SD state in  \Appx{\ref{app: unitary}}.

We will now discuss incorporating the geminal phases with the unitary $\mathcal{J}$ operator.
The geminal phases often provide valuable physical insights; for example, extreme AGPs with specific geminal phase patterns are useful in describing frustrated spin systems 
\cite{Batista2009,Chertkov2021,Pal2021,SpinAGP2023}.
The explicit form of the Jastrow operator $\mathcal{J}$ is  
\begin{equation} \label{eq: ej1}
  \mathcal{J} 
  = e^{i \Jst{1}}
  = \prod_{p = 1}^M \: e^{ i \alpha_p \N{p} / 2 },
\end{equation}
where  
\begin{equation}
  \Jst{1}
  = \frac{1}{2} \: \sum_{p = 1}^M \: \alpha_p \: \N{p}.
\end{equation}
The unitary operator $e^{i \Jst{1}}$ transforms an AGP to another
\begin{equation}
  \rtAGP
  = e^{i \Jst{1}} \: \rAGP,  
\end{equation}
with each geminal coefficient transformed as
$ \tilde{\eta}_p = e^{i \alpha_p} \: \eta_p $ \cite{AGPCC2021,LCAGP2021}.
The operators in \Eq{\ref{eq: ej1}} can be realized by applying parallel $R_z$ gates with $\mathcal{O} (1)$ depth
\begin{equation} \label{eq: rz}
  e^{ i \alpha_p \N{p} / 2 }
  \propto e^{ - i \alpha_p \Sz{p} / 2 } 
  = R_z ( \alpha_p),
\end{equation}
with an inconsequential global phase.
The $R_z ( \alpha_p)$ gate is defined as 
\begin{equation}
  R_z (\alpha_p) 
  = \begin{pmatrix}
  e^{ - i \alpha_p / 2 } & 0 
  \\
  0 & e^{ i \alpha_p / 2 } 
  \end{pmatrix}. 
\end{equation}
Thus, we conclude the AGP preparation algorithm under the paired encoding approach.
We illustrate the complete AGP circuit in \Fig{\ref{fig: fullagp}}. 

The rotation angles of \Eq{\ref{eq: rotagp}} play a central role in our preparation algorithm and they are tailored in such a way that $\Uni{M}{N}$ prepares an AGP, or equivalently, an ESP state. 
However, it is possible to introduce more variational flexibility to $\Uni{M}{N}$ by allowing the $\{ \theta_{pq} \}$ angles to be variationally independent. We explore this idea in \Appx{\ref{app: bts}} which, as we show, will lead to a preparation scheme for a different geminal--based wavefunction that is inspired by AGP. 
We refer to this function as the \textit{binary tree state} and show that it could have a variationally lower energy than AGP by testing it for ground state of the pairing Hamiltonian \cite{Richardson1963,Richardson1964,RG2004}. 

While, in this section, we concentrated on implementing AGP when a set of geminal coefficients are given, it is also possible to use the AGP circuit discussed here to variationally optimize the AGP geminal coefficients in a VQE manner \cite{AAGVQE2014,AAGVQE2016}, should that provide an advantage. 
As noted in \Sec{\ref{sec:intro}}, the geminal coefficients can be optimized efficiently on a classical computer at mean--field cost.

\subsection{Nonzero seniority} \label{subsec:nzero_seniority}

\begin{figure*}[t] 
\centering
\includegraphics[width=0.5\textwidth]{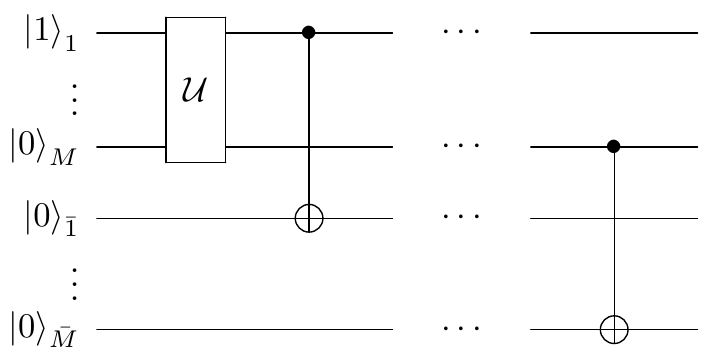}
\caption{
  Circuits to implement AGP for seniority nonzero applications. 
  First, $\mathcal{U} = \mathcal{J} \: \Uni{M}{N}$ is applied to bring the qubits associated with orbitals $\{p\}$ into an ESP state, as discussed in \Sec{\ref{subsec:agp_prep}}.
  Then, a series of CNOT gates are applied to entangle the qubits associated with $\{\bar{p}\}$ with their paired companions, as discussed in \Sec{\ref{subsec:nzero_seniority}}.
}
\label{fig:nonzero_seniority}
\end{figure*}

In this section, we discuss two ways we may extend AGP beyond the seniority--zero space. The first approach relaxes the paired encoding of \Eq{\ref{eqs:pair-qubit-mapping}} and implements AGP in the larger $2M$ qubit space. This approach is one--to--one correspondent to the conventional formulation of AGP in the fermionic space, hence we refer to it as the \textit{fermionic} implementation. The second approach defines AGP directly in the $su(2)$ space, and so, it can be identified as AGP of spin--$1/2$'s or qubits. We refer to this second formulation as \textit{qubit}--AGP. 

\subsubsection{Fermionic implementation} \label{subsec:fermionic-implementation}

Consider the JW transformation as formulated in \Sec{\ref{sec:AGP_as_esp}}. Our goal is to relax the paired encoding for AGP and find a circuit that identically corresponds to its fermionic encoding. To this end, we allocate $2M$ qubits so that half of the qubits represent the ``no-bar" spin orbitals, $\{p\}$, while the other half represents their paired orbital companions, $\{\bar{p}\}$. Our strategy to implement AGP in this space is simple: We use the same $U_{MN}$ circuit introduced in the previous section to bring the qubits associated with orbitals $\{p\}$ into an ESP state 
Then we apply a series of CNOTs targeting qubits $\{\bar{p}\}$ to entangle orbitals $p$ and $\bar{p}$ \cite{Kottmann2022} with the same geminal coefficient. 
Expressed mathematically, we can write
\begin{align} \label{eq:prep_AGP_nonzero_seniority}
    \rAGP &= \prod_{p=1}^M \text{CNOT}_{p,\bar{p}} \OffP{M} \otimes \left(\Uni{M}{N} \OffP{M-N} \OnP{N} \right).
\end{align}
where we assigned the first $M$ qubits to orbitals $\{p\}$ and the second half to $\{\bar{p}\}$ for illustrative simplicity.
\Fig{\ref{fig:nonzero_seniority}} shows the schematic circuit. 

Indeed, the choice for labeling the qubits is arbitrary. 
We could have alternatively chosen to interlace the qubits so that the qubits corresponding to orbitals $p$ and $\bar{p}$ are placed next to each other as in \Reference{\cite{AGPQC2022}}. 
For so doing, we would need to modify the circuit of \Eq{\ref{eq:prep_AGP_nonzero_seniority}} accordingly, or we could append the original circuit with a series of SWAP gates to move the logical qubits to any desired positions.  
If we choose to interlace the qubits, we arrive at a more familiar expression for AGP at the end
\begin{multline} \label{eq:AGPstate_nonzero_seniority}
    \rAGP =
    \frac{1}{\sqrt{ \langle \text{AGP} \rAGP } }
    \SumOp{M}{N} \eta_{p_1} \cdots \eta_{p_N} \\
    \ket{0 \cdots 1_{\bar{p}_N}1_{p_N} \cdots 1_{\bar{p}_2}1_{p_2} \cdots 1_{\bar{p}_1}1_{p_1} \cdots 0}.
\end{multline}
Comparing this with \Eq{\ref{eq:ESPstate}} reveals that the seniority nonzero implementation of AGP can too be viewed as an ESP state, wherein there are $2M$ qubits of which $2N$ pairwise qubits are ``on" in all ${M \choose N}$ possible combinations.

The method presented here can be appended by an orbital rotation operator to further express AGP in other bases. 
This allows for an AGP--optimization scheme on a quantum computer which has been discussed at length in \Reference{\cite{AGPQC2022}}. 

\subsubsection{Qubit--AGP}\label{subsec:qubit-AGP}

There is yet another approach to defining AGP for general Hamiltonians mapped to the $su(2)$ algebra. Instead of aiming to implement AGP that is dual to its fermionic counterpart (i.e. the approach in the preceding section), we make the observation that the mapped Hamiltonian \textit{itself} is a seniority--zero or spin--1/2 Hamiltonian, which allows us to define a \textit{spin}-- or \textit{qubit}--AGP in this space. 

To be precise, consider a generic two--body \textit{ab initio} Hamiltonian
\begin{align}\label{eq:abinitio_Hamiltonian}
    H = \sum_{pq} \: h_{pq} \: \cdag{p} \cn{q} 
    + \frac{1}{4} \: \sum_{pqrs} \: v_{pqrs} \: \cdag{p}\cdag{q} \cn{s}\cn{r},
\end{align}
where $h_{pq}$ and $v_{pqrs}$ are the standard one-- and two--electron integrals, and the indices run from $0$ to $2M$ and are associated with the spin--orbitals. 
After mapping \Eq{\ref{eq:abinitio_Hamiltonian}} to the pairing algebra via the JW transformation, we get a Hamiltonian of the form
\begin{align} \label{eq:abinitio_su2_Hamiltonian}
    H &= \sum_{pq} \: \tilde{h}_{pq} \: \Pdag{p} \Pp{q} \: \hat{\Phi}_{pq} \nonumber
    \\
    &+ \sum_{pqrs} \: \tilde{v}_{pqrs} \: \Pdag{p} \Pdag{q} \Pp{s} \Pp{r} \: \hat{\Phi}_{pqsr},
\end{align}
where $\hat{\Phi}$ contains the appropriate JW strings associated with each term; the relationship between $\tilde{h}_{pq}, \tilde{v}_{pqrs}$ and ${h}_{pq}, {v}_{pqrs}$ have been worked out in \Reference{\cite{TomJWT2022}}. 
Clearly, \Eq{\ref{eq:abinitio_su2_Hamiltonian}} is a seniority--zero Hamiltonian. 
Thus, we can define the qubit-AGP (qAGP) wavefunction as follows:
\begin{multline} \label{eq:qubit-AGP-extended}
    \rqAGP =
    \frac{1}{\sqrt{\langle \text{qAGP} \rqAGP}}
    \SumOp{2M}{2N} \\ \eta_{p_1} \cdots \eta_{p_{2N}} \:
    \ket{0 \cdots 1_{p_{2N}} \cdots 1_{p_2} \cdots 1_{p_1} \cdots 0}.
\end{multline}
where we have $2M$ qubits, each representing the occupation numbers of a spin--orbital. 
There are $2M$ geminal coefficients, and the wavefunction is a superposition of ${2M \choose 2N}$ states. 
In contrast to the approach in the preceding section, there is no pairing scheme defined in \Eq{\ref{eq:qubit-AGP-extended}}, which is easy to see by comparing it to \Eq{\ref{eq:AGPstate_nonzero_seniority}}. 
The BCS wavefunction corresponding to \Eq{\ref{eq:qubit-AGP-extended}}, which we refer to as qubit--BCS (qBCS), can be shown to be 
\begin{multline}\label{eq:bcs_nonzero_seniorty}
    \rqBCS \propto  \prod_{p=1}^{2M} \left(1 + \eta_p \Pdag{p} \right) \rVac \mapsto \\
    \bigotimes_{p=1}^{2M} \: \Big( \cos\left( {\beta_p} \right)\ket{0}_p +  e^{i \alpha_p} \: \sin\left({\beta_p}\right)\ket{1}_p \Big),
\end{multline}
where $\beta_p= \tan^{-1}(|\eta_p|)$. 
We remark that qBCS is the same initial state used in the qubit coupled cluster (QCC) method 
\cite{QubitCC2018,QubitCC2020,QubitCC2021}. 

State preparation of qAGP without number projection can be realized trivially by applying $U_{2M\: 2N}$ to the initial state of $ \OffP{2M-2N} \OnP{2N} $ followed by a layer $\mathcal{J}$ applied to all qubits.


\section{Discussion} \label{sec:discussions}

We have presented a deterministic state preparation algorithm for AGP on a quantum computer. We achieved this by treating AGP as an ESP state, which allowed us to design a divide--and--conquer circuit that generalizes previous research on preparing the Dicke state on a quantum computer. Our circuit can be divided into three broad parts corresponding to implementing 
(i) the magnitudes of geminal coefficients, 
(ii) their phases, and 
(iii) extensions of AGP beyond the seniority--zero space.

The geminal magnitude circuit for AGP can be implemented with a circuit of $\bigO{M^2}$ depth using $\bigO{M}$ qubits. 
We have shown that the corresponding unitary operator is equivalent to a disentangled upCC ansatz with generalized indices whose ordering and amplitudes are tailored to the ESP structure of AGP.
The phases of geminal coefficients are then implemented with parallel $R_z$ gates at $\bigO{1}$ depth, which we have shown to be equivalent to a one--body unitary Jastrow operator in the second quantized formulation. 
Lastly, we discussed two approaches to defining AGP for a generic seniority nonzero space: The first relaxes the paired encoding of the seniority--zero space and is identical to the conventional formulation of AGP in terms of individual fermions. 
We call this approach the fermionic implementation. 
In the second approach, we defined AGP directly in the language of qubits (or spins) for Hamiltonians that get mapped to the $su(2)$ algebra by the JW transformation or similar methods. 
We referred to this second approach as qubit--AGP.
  
In contrast to past work on AGP--based quantum algorithms, where AGP was prepared as PBCS, our preparation scheme has no number projection step. 
This provides a new perspective on preparing symmetry--restored wavefunctions by a unitary evolution on a quantum computer. 
While our work presents one such example for the $U(1)$ gauge symmetry, which includes particle number, it invites the exploration of circuits that prepare other symmetry--restored wavefunctions such as those of $S^2$ or $S_z$ on a quantum computer. 
Similarly, because of the relation between our AGP preparation scheme and disentangled uCC, this work may help design uCC ansatze that include the type of entanglement that originates from symmetry breaking and restoration.

AGP is part of a broader family of geminal product states \cite{AyersJCTC2013,GeminalReview2022}, and state preparations of geminal--based wavefunctions are not straightforward in general.
Our work opens the possibility of efficiently preparing more sophisticated geminal--based states on a quantum computer. In quantum computing and information theory, the Dicke state is known for diverse applications \cite{Childs2002,Ozdemir2007,Prevedel2009,Toth2012,Farhi2014,Ouyang2021,Hadfield2019}.
Since AGP, or equivalently ESP state, is a generalization of the Dicke state, we expect this work to be useful in areas beyond quantum chemistry.

While our primary goal in this paper has been to lay out our algorithm in a concise manner, we are aware that further modifications can be done to reduce the number of CNOTs and tailor the algorithm for specific quantum computing architectures. 
These efforts have recently been explored for the Dicke state \cite{BartschiDicke2019,Mukherjee2020,Aktar2022}, and the authors of \Reference{\cite{Mukherjee2020}} experimented with implementing the Dicke state on a quantum computer. While some of these efforts can be readily extended to our AGP state preparation algorithm, future work could explore further along this direction.

\begin{figure*}[t]
\includegraphics[width=0.9\textwidth]{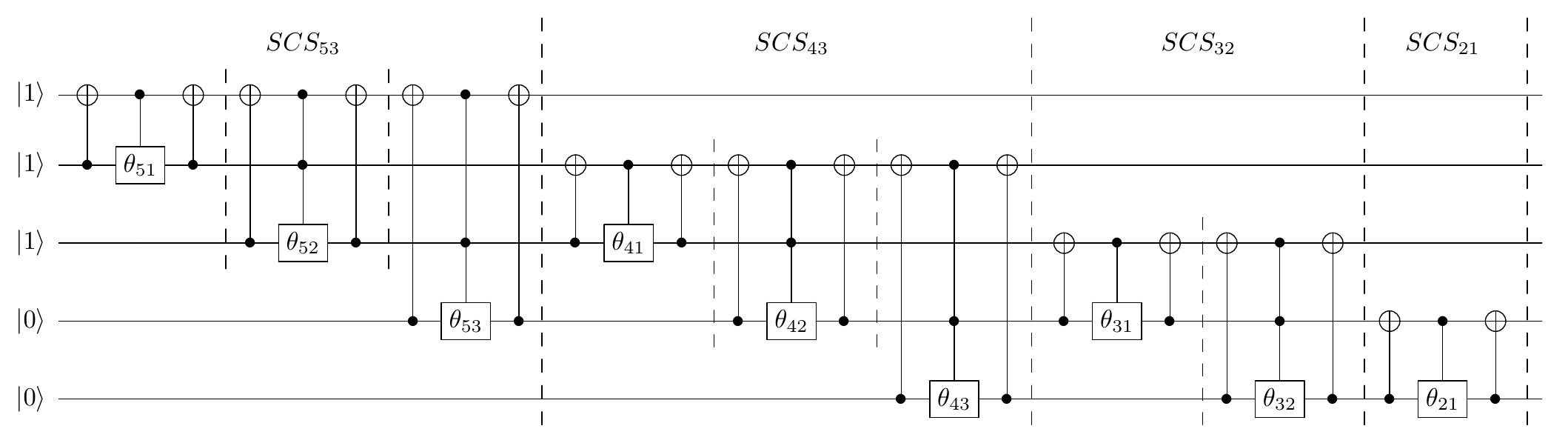}
\caption{
  Circuit for preparing $\Uni{5}{3} \: \ket{00111}$, following \Eq{\ref{eq: evolve}}.
  $\Uni{5}{3}$ is reconstructed in terms of smaller unitary blocks, as discussed in \Sec{\ref{subsec:agp_prep}}.
  Here, each box with $\theta_{p q}$ represents a $R_y (2 \arccos \theta_{p q})$ gate, with the scalars $\theta_{p q}$ defined in \Eq{\ref{eq: rotagp}}. 
}
\label{fig: s53circ}
\end{figure*}


\section{Acknowledgments}

This work was supported by the U.S. Department of Energy under Award No. DE-SC0019374. 
G.E.S. is a Welch Foundation Chair (C-0036). 


\appendix

\section{Unitary evolution example} \label{app: unitary}

We will illustrate preparing $\Uni{M}{N} \: \rHF$ in the paired encoding approach with an example by choosing $M = 5$ and $N = 3$.
First, the operator $\Uni{5}{3}$ is constructed according to \Eq{\ref{eq: unitary}}, also illustrated in \Fig{\ref{fig: block}}.
Then the various $SCS$ operators are constructed according to \Eq{\ref{eq:scs_decomp}}.
The entire circuit for $ \Uni{5}{3} \: \ket{00111} $ is shown in \Fig{\ref{fig: s53circ}} and provides an opportunity for the reader to compare it with the circuit illustration for Dicke state $\rDicke{5}{3}$ in \Reference{\cite{BartschiDicke2019}}.
Following \Fig{\ref{fig: s53circ}}, we get the following states at each step 
\begin{subequations} \label{eq:s53_terms}
\begin{align} 
  &\SCS{5}{3} \: \ket{00111}    
  = \theta_{5 3} \ket{00111} 
  + \bar{\theta}_{5 3} \ket{01110}, 
  \\ \nonumber \\ 
  &\SCS{4}{3} \: \SCS{5}{3} \: \ket{00111}
  = \theta_{4 2} \theta_{5 3} \ket{00111} 
  + \bar{\theta}_{4 2} \theta_{5 3} \ket{01101} \nonumber 
  \\
  &\hskip6ex
  + \theta_{4 3} \bar{\theta}_{5 3} \ket{01110} 
  + \bar{\theta}_{4 3} \bar{\theta}_{5 3} \ket{11100},
  \\ \nonumber \\
  &\SCS{3}{2} \: \SCS{4}{3} \: \SCS{5}{3} \: \ket{00111}
  = \theta_{3 1} \theta_{4 2} \theta_{5 3} \ket{00111} \nonumber 
  \\
  &\hskip6ex
  + \bar{\theta}_{3 1} \theta_{4 2} \theta_{5 3} \ket{01011} 
  + \theta_{3 2} \bar{\theta}_{4 2} \theta_{5 3} \ket{01101} \nonumber 
  \\
  &\hskip6ex
  + \bar{\theta}_{3 2} \bar{\theta}_{4 2} \theta_{5 3} \ket{11001} 
  + \theta_{3 2} \theta_{4 3} \bar{\theta}_{5 3} \ket{01110} \nonumber 
  \\
  &\hskip6ex
  + \bar{\theta}_{3 2} \theta_{4 3} \bar{\theta}_{5 3} \ket{11010} 
  + \bar{\theta}_{4 3} \bar{\theta}_{5 3} \ket{11100},
  \\ \nonumber \\ 
  &\SCS{2}{1} \: \SCS{3}{2} \: \SCS{4}{3} \: \SCS{5}{3} \: \ket{00111}
  = \theta_{3 1} \theta_{4 2} \theta_{5 3} \: \ket{00111} \nonumber 
  \\
  &\hskip6ex
  + \theta_{2 1} \bar{\theta}_{3 1} \theta_{4 2} \theta_{5 3} \: \ket{01011} 
  +  \bar{\theta}_{2 1} \bar{\theta}_{3 1} \theta_{4 2} \theta_{5 3} \: \ket{10011} \nonumber 
  \\
  &\hskip6ex
  + \theta_{2 1} \theta_{3 2} \bar{\theta}_{4 2} \theta_{5 3} \: \ket{01101} 
  + \bar{\theta}_{2 1} \theta_{3 2} \bar{\theta}_{4 2} \theta_{5 3} \: \ket{10101} \nonumber 
  \\
  &\hskip6ex
  + \bar{\theta}_{3 2} \bar{\theta}_{4 2} \theta_{5 3} \: \ket{11001} 
  + \theta_{2 1} \theta_{3 2} \theta_{4 3} \bar{\theta}_{5 3} \: \ket{01110} \nonumber 
  \\
  &\hskip6ex
  + \bar{\theta}_{2 1} \theta_{3 2} \theta_{4 3} \bar{\theta}_{5 3} \: \ket{10110} 
  + \bar{\theta}_{3 2} \theta_{4 3} \bar{\theta}_{5 3} \: \ket{11010} \nonumber 
  \\
  &\hskip6ex
  + \bar{\theta}_{4 3} \bar{\theta}_{5 3} \: \ket{11100},
\end{align}
\end{subequations} 
where $\theta_{p q}$ is defined in \Eq{\ref{eq: rotagp}} and 
\begin{equation}
  \bar{\theta}_{p q} 
  \equiv \sqrt{1 - \theta_{p q}^2} 
  = \Nfac{p-1}{q}{p}{q}.
\end{equation}
Simplifying \Eq{\ref{eq:s53_terms}} leads to 
\begin{align} \label{eq:s53_terms_simplified}
  \Uni{5}{3} \: \ket{00111}    
  &= \frac{1}{\sqrt{\ESP{5}{3}}} \: \Big(
  | \eta_1 \eta_2 \eta_3 | \: \ket{00111} 
  + | \eta_1 \eta_2 \eta_4 | \: \ket{01011} \nonumber 
  \\
  &+ \sqrt{\ESP{1}{1}} \: | \eta_1 \eta_2 | \: \ket{10011} 
  + | \eta_1 \eta_3 \eta_4 | \: \ket{01101} \nonumber 
  \\
  &+ \sqrt{\ESP{1}{1}} \: | \eta_1 \eta_3 | \: \ket{10101} 
  + \sqrt{\ESP{2}{2}} \: | \eta_1 | \: \ket{11001} \nonumber 
  \\
  &+ | \eta_2 \eta_3 \eta_4 | \: \ket{01110} 
  + \sqrt{\ESP{1}{1}} \: | \eta_2 \eta_3 | \: \ket{10110} \nonumber 
  \\
  &+ \sqrt{\ESP{2}{2}} \: | \eta_2 | \: \ket{11010} 
  + \sqrt{\ESP{3}{3}} \: \ket{11100}
  \Big).
\end{align}
As discussed in \Sec{\ref{subsec:agp_prep}, $\Uni{M}{N}$ implements the magnitudes of the AGP geminal coefficients. 
Thus, we can simplify a few terms of \Eq{\ref{eq:s53_terms_simplified}} further as }
$ \sqrt{\ESP{1}{1}} = | \eta_5 | $,  
$ \sqrt{\ESP{2}{2}} = | \eta_4 \eta_5 | $, and 
$ \sqrt{\ESP{3}{3}} = | \eta_3 \eta_4 \eta_5 | $.
Finally, the right hand side of \Eq{\ref{eq:s53_terms_simplified}} turns into the $\rESP{5}{3}$ state
\begin{align} 
  \rESP{5}{3}   
  &= \frac{1}{\sqrt{\ESP{5}{3}}} \: \Big(
  | \eta_1 \eta_2 \eta_3 | \: \ket{00111} 
  + | \eta_1 \eta_2 \eta_4 | \: \ket{01011} \nonumber 
  \\
  &+ | \eta_1 \eta_2 \eta_5 | \: \ket{10011} 
  + | \eta_1 \eta_3 \eta_4 | \: \ket{01101} \nonumber 
  \\
  &+ | \eta_1 \eta_3 \eta_5 |  \: \ket{10101} 
  + | \eta_1 \eta_4 \eta_5 |  \: \ket{11001} \nonumber 
  \\
  &+ | \eta_2 \eta_3 \eta_4 | \: \ket{01110} 
  + | \eta_2 \eta_3 \eta_5 | \: \ket{10110} \nonumber 
  \\
  &+ | \eta_2 \eta_4 \eta_5 | \: \ket{11010} 
  + | \eta_3 \eta_4 \eta_5 | \: \ket{11100}
  \Big),
\end{align}
with the geminal coefficient magnitudes. 
The corresponding phases can be added by the unitary operator $\mathcal{J}$, as discussed in \Sec{\ref{subsec:agp_prep}}.


\section{New polynomial state} \label{app: bts}


\begin{algorithm}[b]
  \SetAlgoNoLine
  \caption{ 
    Computation of \Eq{\ref{eq:norm_bts}}. } \label{alg:sumbtp}
  \KwIn{ $\mathbf{x}_{N \times M}$ matrix with $x_p^j = | \eta_p^j |^2$ for all $j$ and $p$.}
  \KwOut{$ \BTP{M}{N} \left( \textbf{x} \right) = \BTP{M}{N} $.}
  function $ \BTP{M}{N} = \texttt{NormBTS} \left( \textbf{x} \right)$ 
  \\
  $ \BTP{p}{0} = 1, \quad 1 \leq p \leq M - 1; $ 
  \\
  $ \BTP{p}{q} = 0, \quad q > p; $
  \\
  $ \BTP{1}{1} = x_1^1; $
  \\
  for $p = 2:M$
  \\
  \hskip3ex for $q = max(1, p + N - M):min(p, N)$
  \\
  \hskip6ex 
  $ \BTP{p}{q} 
  = x_p^q \: \BTP{p - 1}{q - 1} + \BTP{p - 1}{q} $; 
  \\ 
  \hskip3ex end 
  \\
  end
\end{algorithm}


As discussed in \Sec{\ref{subsec:agp_prep}}, state preparation of AGP could inspire the preparation of other geminal--based wavefunctions. 
Perhaps the simplest way to go beyond AGP is to allow the rotation angles of the circuit to vary independently, thereby adding more variational flexibility to the reference state. 
Note that the rotation angles discussed in \Sec{\ref{subsec:agp_prep}}, $\{\theta_{pq}\}$, are interdependent as they are engineered in a way to implement AGP. 
This can be easily verified by observing that there are $\bigO{M}$ geminal coefficients, but there are $\bigO{M^2}$ rotation angles in the circuit.
 
Let $\{ \tilde{\theta}_{pq} \}$ to be the set of the new rotation angles which we would like to vary independently. 
By substituting $\theta_{pq} \xrightarrow{} \tilde{\theta}_{pq}$ in the circuit and taking into account the normalization of the wavefunction at every splitting, we can analytically work out what will be the resulting wavefunction. 
We refer to this wavefunction as a \textit{binary tree state} (BTS) for reasons that will become apparent shortly. 

Define $\rBTS{M}{N}$ to be a BTS with $N$ pairs and $M$ paired orbitals ($1 \leq N \leq M$) over a set of two--tensor elements $\{\eta_p^{j}\}$ as follows 
\begin{multline} \label{eq:BTS}
   \rBTS{M}{N} \equiv 
   \frac{1}{\sqrt{\BTP{M}{N}}} 
   \SumOp{M}{N} \eta_{p_1}^1 \cdots \eta_{p_N}^N \\
   \ket{0 \cdots 1_{p_N} \cdots 1_{p_2} \cdots 1_{p_1} \cdots 0}. 
\end{multline}
where we seek $\eta_p^{j}$ of the form $\eta_p^{j} = |\eta_{p}^j| \: e^{i \alpha_p}$ 
such that $\eta_p^{j}$ is nonzero
when $ \text{max} (1, p + N - M) \leq j \leq \text{min} (p, N) $.
The normalization factor, $1/\sqrt{\BTP{M}{N}}$, can be obtained from
\begin{equation} \label{eq:norm_bts}
  \BTP{M}{N}   
  = \SumOp{M}{N} | \eta_{p_1}^1 |^2 \cdots | \eta_{p_N}^N |^2.
\end{equation}
which we refer to as a \textit{binary tree polynomial} (BTP). 
\Eq{\ref{eq:norm_bts}} can be computed efficiently with a little modification to the \texttt{SumESP} algorithm discussed in \Reference{\cite{Jiang2016}}; we refer to \Alg{\ref{alg:sumbtp}} for more details.

Just as in AGP, the BTS $\rBTS{M}{N}$ has the following recursion relation 
\begin{align} \label{eq: recbts}
   \rBTS{p}{q}
   &= \NfacBTP{p - 1}{q - 1}{p}{q} \: \eta_{M - p + 1}^{N - q + 1} \: \rBTS{p - 1}{q - 1} \rOn \nonumber 
   \\
   &+ \NfacBTP{p - 1}{q}{p}{q} \: \rBTS{p - 1}{q} \rOff
\end{align}
for any given $1 \leq q  \leq N < p \leq M$. Written as such, the relationship between $\{\tilde{\theta}_{pq}\}$ and $\{|\eta_p^{j}|\}$ is easy to infer; 
it can be shown that the following rotation angles 
\begin{equation} \label{eq:rot_bts}
   \tilde{\theta}_{pq} 
   = | \eta_{M - p + 1}^{N - q + 1} | \: \NfacBTP{p - 1}{q - 1}{p}{q}
\end{equation}
in the unitary gate $\mathcal{U}$ of \Sec{\ref{subsec:agp_prep}} prepares $\rBTS{M}{N}$ with coefficients 
$\eta_p^j = |\eta_p^j| \: e^{i \alpha_p}$. The AGP and BTS recursions follow a binary tree structure, hence the name for \Eq{\ref{eq:BTS}}.

When considering the seniority--zero space, both BTS and AGP approximate the doubly occupied configuration interaction (DOCI) wavefunction \cite{DOCI1967,Seniority2011}
\begin{multline} \label{eq:doci}
  \ket{\text{DOCI}} \equiv 
  \SumOp{M}{N} D_{p_1 \cdots p_N} \\
  \ket{0 \cdots 1_{p_N} \cdots 1_{p_2} \cdots 1_{p_1} \cdots 0},
\end{multline}
with a monomial decomposition of the DOCI coefficients, i.e. $\{ D_{p_1 \cdots p_n} \}$.
DOCI is the most general seniority--zero state, and by comparing \Eq{\ref{eq:ESPstate}}, \Eq{\ref{eq:BTS}}, and \Eq{\ref{eq:doci}}, it is clear that BTS allows more flexibility into approximating DOCI than AGP.
Indeed, BTS reduces to AGP when $\eta_p^j = \eta_p$ for all $j$.


\begin{figure}[t]
\includegraphics[width=0.9\columnwidth]{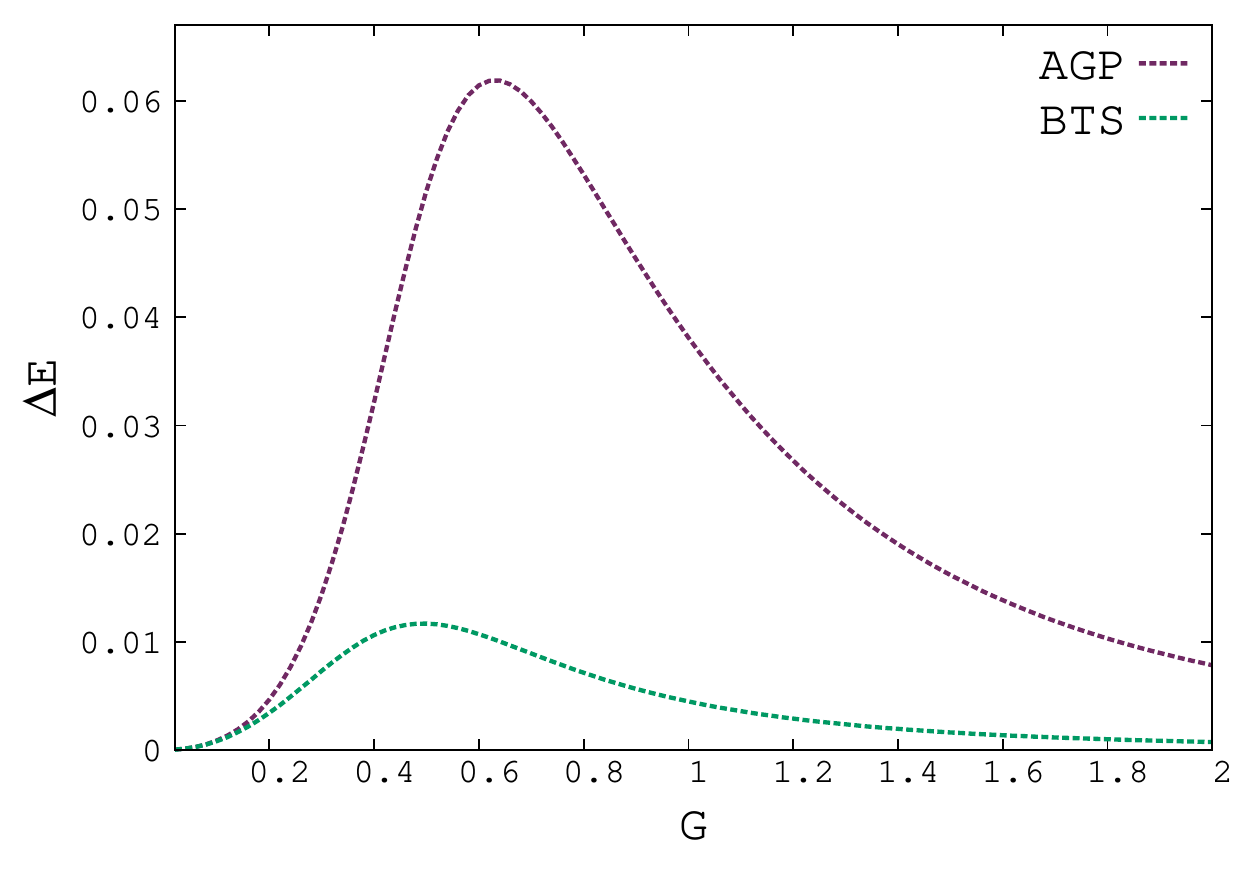}
\caption{
  Total energy errors ($E_{\text{method}} - E_{\text{exact}}$) for the reduced BCS Hamiltonian with respect to its two-body interaction parameter $G$, as defined in \Eq{\ref{eq:RBCS_ham}}.
  The number of paired orbitals is $M = 10$, and the number of fermion pairs is $N = 5$.
  The methods are the variationally best single AGP of \Eq{\ref{eq: gemagp}} and BTS of \Eq{\ref{eq:BTS}}.
}
\label{fig: EnM10N8}
\end{figure}


We can put this idea to the test by comparing the variationally best energies of the two wavefunctions for a model Hamiltonian. The exactly solvable pairing Hamiltonian \cite{Richardson1963,Richardson1964,RG2004} is suitable for this purpose
\begin{equation} \label{eq:RBCS_ham}
  \Ham 
  = \sum_p \: \epsilon_p \: \N{p}
  - G \: \sum_{pq} \: \Pdag{p} \Pp{q}, 
\end{equation}
where the one-body interaction is $ \epsilon_p = p$, and the scalar $G$ tunes the strength of pairwise interactions.
We compare the total energy errors of BTS and AGP over a range of $G$ values in \Fig{\ref{fig: EnM10N8}} by minimizing the energy using an in--house code. 
Results show that BTS energies are lower than or comparable to those of AGP in all correlation regimes, which supports our hypothesis that BTS could be a more accurate wavefunction than AGP due to its variational flexibility.


\bibliography{AGP}

\end{document}